\documentclass[authoryear,preprint,12pt]{style_}

\usepackage{graphicx}
\usepackage{amssymb}
\begin{document}

\begin{frontmatter}


% use the tnoteref command within \title for footnotes;
% use the tnotetext command for theassociated footnote;
% use the fnref command within \author or \address for footnotes;
% use the fntext command for theassociated footnote;
% use the corref command within \author for corresponding author footnotes;
% use the cortext command for theassociated footnote;
% \author{Name\corref{cor1}\fnref{label2}}
% \address{Address\fnref{label3}}

\title{High Order Phase-fitted Discrete Lagrangian Integrators for Orbital Problems}


\author[uop]{O.T. Kosmas}
\ead{odykosm@uop.gr}

\author[uop]{D.S. Vlachos\corref{cor}}
\ead{dvlachos@uop.gr}

\cortext[cor]{Corresponding author}
\address[uop]{Department of Computer Science and Technology,\\
Faculty of Sciences and Technology, University of Peloponnese\\
GR-22 100 Tripolis, Terma Karaiskaki, GREECE}

\begin{abstract}
% Text of abstract
In this work, the benefits of the phase fitting technique are embedded in high order discrete Lagrangian integrators. The proposed methodology creates integrators with zero phase lag in a test Lagrangian in a similar way used in phase fitted numerical methods for ordinary differential equations. Moreover, an efficient method for frequency evaluation is proposed based on the eccentricities of the moving objects. The results show that the new method dramatically improves the accuracy and total energy behaviour in Hamiltonian systems. Numerical tests for the 2-body problem with ultra high eccentricity up to $0.99$ for $10^6$ periods and to the Henon-Heiles Hamiltonian system with chaotic behaviour, show the efficiency of the proposed approach.\end{abstract}

\begin{keyword}
Phase Fitting \sep Exponential Fitting \sep Discrete Lagrangian Integrators

\PACS 02.60,Jh \sep 45.10.-b \sep 45.10.Db \sep 45.10.Hj \sep 45.10.Jf

% MSC codes here, in the form: \MSC code \sep code
% or \MSC[2008] code \sep code (2000 is the default)

\end{keyword}

\end{frontmatter}

% main text
\section{Introduction}
\label{section_intro}
In the field of numerical integration, methods specially tuned on oscillating functions, are of great practical importance. Such methods are needed in various branches of natural sciences, particularly in physics, since a lot of physical phenomena exhibit a pronounced oscillatory behaviour. For a review of such methods see \cite{ixaru_CPC_100_56_70,vandenberghe_CPC_123_7_15,vandenberghe_CPC_150_346_357,ixaru_CPC_150_116_128,vandaele_APNUM_57_415_435} and references there in as well as the book \cite{ixaru_Book_EF_KAP_04}.

For problems having highly oscillatory solutions, standard methods with unspecialized use can require a huge number of steps to track the oscillations. One way to obtain a more efficient integration process is to construct numerical methods with an increased algebraic order, although the simple implementation of high algebraic order methods may cause several problems (for example, the existence of parasitic solutions \cite{quinlan_arxiv_astro_ph_9901136}). On the other hand, there are some special techniques for optimizing numerical methods. Trigonometrical fitting and phase-fitting are some of them, producing methods with variable coefficients, which depend on $v = \omega h$, where $\omega$ is the dominant frequency of the problem and $h$ is the step length of integration. This technique is known as exponential (or trigonometric if $\mu=i\omega$) fitting and has a long history \cite{gautschi_NM_3_381_61}, \cite{lyche_NM_19_65_72}. An important property of exponential fitted algorithms is that they tend to the classical ones when the involved frequencies tend to zero, a fact which allows to say that exponential fitting represents a natural extension of the classical polynomial fitting. The examination of the convergence of exponential fitted multistep methods is included in Lyche’s theory \cite{lyche_NM_19_65_72}. The general theory is presented in detail in \cite{ixaru_Book_EF_KAP_04}. Furthermore, considering the accuracy of a method when solving oscillatory problems, it is more appropriate to work with the phase-lag, rather than its usual primary local truncation error. We mention the pioneering paper of Brusa and Nigro \cite{brusa_IJNME_15_685_80}, in which the phase-lag property was introduced. This is actually another type of a truncation error, i.e. the angle between the analytical solution and the numerical solution. A significant application of the phase or exponential fitting is on the construction of symplectic methods for oscillatory problems encountered in physics and chemistry (\cite{monovasilis_JMC_37_3263_05,monovasilis_JMC_40_3257_06}). 

Although phase fitting and exponential fitting are a major improvement over algebraic fitted methods especially for oscillatory and orbital problems, there is not significant evidence from published results that these methods can be applied for long term integration (for example for millions or billions of periods). Moreover, several authors use to test their methods to the well known 2-body problem but only for relatively low eccentricities (up to $0.2$) and for relatively small number of periods (no more than several thousands). We mention here the efforts of \cite{vyver_NA_11_8_577_06,vyver_NA_10_4_261_05,wang_NA_11_2_90_05,simos_NA_9_1_59_04,anastassi_NA_10_4_301_05,anastassi_NA_10_1_31_04} in which there is no evidence that the phase fitting or trigonometric fitting can be applied to high eccentricities (for example to the Halley comet with eccentricity close to $0.967$) and for long time.

Another approach to oscillatory and especially Hamiltonian systems is the theory of discrete variational mechanics, which was set up in the 1960s \cite{jordan_JEL_17_697_64,cadzow_IJC_11_393_70,logan_AM_9_210_73} and then it was proposed in the optimal control literature. It then motivated a lot of authors and soon the discrete Euler-Lagrange equations were formulated and the first integrators in the discrete calculus of variation and further the multi-freedom and higher-order problems were studied. Afterwards, the canonical structure and symmetries for discrete systems were obtained, and Noether's theorem to the discrete case was extended \cite{maeda_MJ_25_405_80,maeda_MJ_26_85_81}. Finally, the time as a discrete dynamical variable was regarded \cite{lee_PLB_122_217_83}. A detailed description of the essential properties of variational integrators can be found in \cite{marsden_AN_10_357_01,marsden_CMP_199_351_98,lee_PLB_122_217_83}. One of the most important properties of variational integrators is that since the discrete Lagrangian is an approximation of a continuous Lagrangian function, the obtained numerical integrator inherits some of the geometric properties of the continuous Lagrangian (such as symplecticity, momentum preservation).

In the present work, the benefits of the two approach are combined in order to construct high order discrete Lagrangian integrators with phase fitting. To obtain this, we have adopted a test Lagrangian problem (similar to the test ODE in the phase fitting) which is the harmonic oscillator with given frequency $\omega$. Then, we construct discrete variational schemes that solve exactly the test Lagrangian. The application of the method to a general Lagrangian needs the determination of the frequency $\omega$ at every step of the integration. The method is applied to the 2-body problem with eccentricity up to $0.99$ for $10^6$ periods and to Henon-Heiles system which for high energies exhibit chaotic behaviour. The results clearly demonstrate the efficiency of the new approach.

\section{Discrete Variational Mechanics}
\label{section_DVM}
The well known least action principle of the continuous Lagrange - Hamilton Dynamics can be used as a guiding principle to derive discrete integrators. Following the steps of the derivation of Euler-Lagrange equations in the continuous time Lagrangian dynamics, one can derive the discrete time Euler-Lagrange equations. For this purpose, one considers positions $q_{0}$ and $q_{1}$ and a time step $h\in{R}$, in order to replace the parameters of position $q$ and velocity $\dot{q}$ in the continuous time Lagrangian $L (q, \dot{q}, t)$. Then, by considering the variable $h$ as a very small (positive) number, the positions $q_{0}$ and $q_{1}$ could be thought of as being two points on a curve (trajectory of the mechanical system) at time $h$ apart. Under these assumptions, the following approximations hold: $$q_{0}\approx q(0)\, , \qquad \qquad q_{1}\approx q(h) \, ,$$ and a function $L_{d}(q_{0},q_{1},h)$ could be defined known as a discrete Lagrangian function.

Many authors assume such functions to approximate the action integral along the curve segment between $q_{0}$ and $q_{1}$, i.e.
\begin{equation}
 L_{d}(q_{0},q_{1},h)=\int_{0}^{h}L(q,\dot{q},t) dt
\end{equation}
Furthermore, one may consider the very simple approximation for this integral given on the basis of the rectangle rule described in \cite{marsden_AN_10_357_01}. According to this rule, the integral $\int_{0}^{T}{Ldt}$ could be approximated by the product of the time-interval ${h}$ times the value of the integrand $L$ obtained with the velocity $\dot{q}$ replaced by the approximation $(q_{1}-q_{0})/h$:  The next step is to consider a discrete curve defined by the set of points $\{q_{k}\} _{k=0}^{N}$, and calculate the discrete action along this sequence by summing the discrete Lagrangian of the form $L_{d}(q_{k},q_{k+1},h)$ defined for each adjacent pair of points $(q_{k}$, $q_{k+1})$. 

Following the case of the continuous dynamics, we compute variations of this action sum with the boundary points $q_{0}$ and $q_{N}$ held fixed. Briefly, discretization of the action functional leads to the concept of an action sum 
\begin{equation}
S_{d}(\gamma_{d})=\sum_{k=1}^{n-1}L_{d}(q_{k-1},q_{k}),
\qquad \gamma_{d}=(q_{0},...,q_{n-1})\in Q^{n}
\end{equation}
where $L_{d} : Q \times Q \rightarrow R$ is an approximation of L called the discrete Lagrangian. Hence, in the discrete setting the correspondence to the velocity phase space $TQ$ is $Q \times Q$. An intuitive motivation for this is that two points close to each other correspond approximately to the same information as one point and a velocity vector. The discrete Hamilton's principle states that if $\gamma_{d}$ is a motion of the discrete mechanical system then it extremizes the action sum, i. e., $\delta S_{d}=0$. By differentiation and rearranging of the terms and having in mind that both $q_0$ and $q_N$ are fixed, the discrete Euler-Lagrange (DEL) equation is obtained:
\begin{equation}
\label{equ_DEL}
D_{2}L_{d}(q_{k-1},q_{k},h)+D_{1}L_{d}(q_{k},q_{k+1},h)=0
\end{equation}
where the notation $D_{i}L_{d}$ indicates the slot derivative with respect to the argument of $L_{d}$.

We can define now the map $\Phi:Q\times Q\rightarrow Q\times Q$, where $Q$ is the space of generalized positions $q$, by which
\begin{equation}
 D_1L_d\circ \Phi+D_2L_d=0
\end{equation}
which means that $\Phi (q_{k-1},q_k)=(q_k,q_{k+1})$. Then, if for each $q\in Q$, the map $D_1L_d(q,q):T_qQ\rightarrow T^*_qQ$ is invertible, then $D_1L_d: Q\times Q\rightarrow T^*Q$ is locally invertible and so the discrete flow defined by the map $\Phi$ is well defined for small enough time steps (see \cite{kane_JMP_40_7_3353_99} for details). Moreover, if we define the fiber derivative
\begin{equation}
 FL_d:Q\times Q\rightarrow T^*Q
\end{equation}
and the two-form $\omega$ on $Q\times Q$ by pulling back the canonical two-form $\Omega_{CAN}=dq^i\wedge dp_i$ from $T^*Q$ to $Q\times Q$:
\begin{equation}
 \omega=FL^*_d(\Omega_{CAN})
\end{equation}
The coordinate expression for $\omega$ is
\begin{equation}
 \omega=\frac{\partial ^2L_d}{\partial q^i_k \partial q^j_{k+1}}(q_k,q_{k+1})dq^i_k\wedge d^j_{k+1}
\end{equation}
and can be easily proved that the map $\Phi$ preserves the symplectic form $\omega$ (two different proofs are presented in \cite{marsden_CMP_199_351_98} and \cite{wendlandt_PD_106_223_97}). Finally, assuming that the discrete Lagrangian is invariant under the action of a Lie group $G$ on $Q$ and $\xi \in g$, the Lie algebra of $G$, by analogy with the continuous case, we can define the discrete momentum map $J_d:Q\times Q\rightarrow g^*$ by
\begin{equation}
 \left<J_d(q_k,q_{k+1}),\xi\right>:=\left<D_aL_d(q_k,q_{k+1},\xi_Q(q_k)\right>
\end{equation}
It can be proved that the map $\Phi$ preserves the momentum map $J_d$ \cite{wendlandt_PD_106_223_97}.

In a position-momentum form the discrete Euler-Lagrange equations (\ref{equ_DEL}) can be defined by the equations below 
\begin{eqnarray}
\label{equ_DHP}
\nonumber p_{k}&=&-D_{1}L_{d}(q_{k},q_{k+1},h) \\
p_{k+1}&=&D_{2}L_{d}(q_{k},q_{k+1},h)
\end{eqnarray}

\section{Phase-fitted Discrete Lagrangian Integrators}
Summarizing the phase fitting technique, we consider for simplicity only first order differential equations, although the same results can be easily obtained for second order equations too. Consider the test problem 
\begin{equation}
\frac{dy(t)}{dt}=i\omega _0 y(t),\;y(0)=1 
\label{equ_test_equ}
\end{equation}
with exact solution 
\begin{equation}
y(t)=e^{i\omega _0 t}
\end{equation}
where $\omega _0$ is a non-negative real value. Let $\hat{\Phi}(h)$ be a numerical map which when it is applied to a set of known past values, it produces a numerical estimation of $y(t+h)$. If we assume that all past values are known exactly, then the numerical estimation $\hat{y}(t+h)$ of $y(t+h)$ will be
\begin{equation}
\hat{y}(t+h)=\alpha (\omega _0 h) \cdot e^{i(\omega _0 t+\phi (\omega _0 h) )}
\end{equation}
while the exact solution is $e^{i (\omega _0 t +\omega _0 h)}$. Then the ratio of the estimated to the exact solution is
\begin{equation}
L=\frac{\hat{y}(t+h)}{y(t+h)}=\alpha (\omega _0 h)e^{-i(\omega _0 h-\phi (\omega
_0 h) ) }
\label{equ_def_PL}
\end{equation}
In the above equation (\ref{equ_def_PL}), the term $\alpha (\omega _0 h)$ is called the \textit{amplification error}, while the term $l(\omega _0 h)=\omega
_0 h-\phi(\omega _0 h)$ is called the \textit{phase lag} of the numerical map. In the case that $\alpha (\omega _0 h)=1$ and $l(\omega _0 h)=0$, we say that
the numerical map $\Phi(h)$ is \textit{exponentially fitted} at the frequency $\omega _0$ and at the step size $h$. The technique of phase fitting can now be considered as the vanishing or minimization of the phase lag.

Consider now the discrete Lagrangian $L_d(q_k,q_{k+1},h)$ ($q_k$ corresponds to time $t_k$ and $q_{k+1}$ to time $t_{k+1}=t_k+h$) and a set of $s$ intermediate points $q^j$ with $q^j=q\left(t_k+c^jh\right)$. The role of the number of intermidiate points will be discussed later. Assuming that $c^1=0$ and $c^s=1$ we always have $q^1=q_k$ and $q^s=q_{k+1}$. Then we can approximate $L_d$ with the quadrature
\begin{equation}
 L_d(q_k,q_{k+1},h)=h\sum_{j=1}^s w_j\cdot L\left( q(t_k+c^jh),\dot{q}(t_k+c^jh),c^jh \right)
\end{equation}
For maximal algebraic order it is easily proved that the following conditions must hold:
\begin{equation}
 \sum_{j=1}^s w_j\left( c^j \right) ^l=\frac{1}{l+1}\;,\;l=0,1,..
\end{equation}
Then, we can approximate intermediate points and their derivatives with
\begin{equation}
\label{equ_inter}
 \begin{array}{l}
  q^j=b^jq_k+\bar{b}^jq_{k+1} \\
  \dot{q}^j=\frac{1}{h}\left( B^jq_k+\bar{B}^jq_{k+1}\right)
 \end{array}
\end{equation}
Consider now the test Lagrangian (harmonic oscillator) similar to the test equation (\ref{equ_test_equ})
\begin{equation}
\label{equ_Ltest}
 L_t=\frac{1}{2}\dot{q}^2-\frac{1}{2}\omega ^2 q^2
\end{equation}
Then, applying the above assumptions in Eq. (\ref{equ_DEL}) we get
\begin{equation}
 q_{k+1}=\frac{\sum_{j=1}^sw_j\left( \left(B^j\right)^2+\left(\bar{B}^j\right)^2-u^2\left(\left(b^j\right)^2+\left(\bar{b}^j\right)^2\right)\right)}{\sum_{j=1}^sw_j\left(u^2b^j\bar{b}^j-B^j\bar{B}^j\right)}q_k-q_{k-1}
\end{equation}
where $u=\omega h$. Since the exact solution of Eq. (\ref{equ_Ltest}) is
\begin{equation}
 q(t)=Ae^{i\omega t}+Be^{-i\omega t}
\end{equation}
and we want to force our method to solve exactly Eq. (\ref{equ_DEL}), we get
\begin{eqnarray}
\nonumber   b^j&=&cos \left( c^ju \right)-\frac{cosu}{sinu}sin \left( c^ju \right) \\
\nonumber \bar{b}^j&=&\frac{ sin \left( c^ju \right) }{sinu} \\
\nonumber B^j&=&-usin\left( c^ju \right)-u\frac{cosu}{sinu}cos\left( c^ju \right)\\
\bar{B}^j&=&u\frac{cos\left( c^ju \right) }{sinu}
\end{eqnarray}

The role of intermediate points will be clarified now by introducing corrections at the these points. Let $x_j,j=1,\ldots,S$ a set of parameters and rewrite eq. (\ref{equ_inter}) as
\begin{equation}
 \begin{array}{l}
  q^j=b^jq_k+\bar{b}^j\left( x_S-x_{S-1} \right) +p(x_1,x_2,\ldots ,x_{S-1},t)\\
  \dot{q}^j=\frac{1}{h}\left( B^jq_k+\bar{B}^j\left( x_S-x_{S-1} \right)\right)+\dot{p}(x_1,x_2,\ldots ,x_{S-1},t)
 \end{array}
\end{equation}
where the $p(x_1,x_2,\ldots ,x_{S-1},t)$ is the interpolating polynomial of the set of points $\{(t_k,0),(t_k+c^2h,x_1),(t_k+c^3h,x_2),\ldots ,(t_{k+1},x_{S-1})\}$ and $\dot{p}$ its time derivative. Thus, $x_1,x_2,\ldots ,x_{S-1}$ can be considered as corrections to points $q^2,q^3,\ldots ,q^S$ (the correction at $q^1$ is zero since we have assumed that $q^1=q_k$). The set of equations (\ref{equ_DHP}) are now rewritten as
\begin{eqnarray}
\frac{\partial L_d\left(q_k,x_1,x_2,\ldots ,x_S,h\right)}{\partial q_k}+\lambda (x_S-q_{k+1})=0 \nonumber \\
\frac{\partial L_d\left(q_k,x_1,x_2,\ldots ,x_S,h\right)}{\partial x_j}=0\;,\;j=1,2,\ldots ,S-1 \nonumber \\
p_{k+1}=\frac{\partial L_d\left(q_k,x_1,x_2,\ldots ,x_S,h\right)}{\partial x_S}
\end{eqnarray}
where $\lambda$ is the Lagrange multiplier and can be easily proved that it is equal to $p_k$. This technique is similar to those described in \cite{leok_arXiv_math_0508360_05} and \cite{kharevych_AGM_SIGGRAPH_06}.

\section{Frequency Evaluation}
In order to efficiently evaluate the frequency of the problem, we focus on orbital problems and especially on the eccentricity. In general, an elliptic orbit may be parameterized as
\begin{eqnarray}
 x=x_m\cdot cosu +c_1 \nonumber \\
y=y_m\cdot sinu+c_2 \nonumber
\end{eqnarray}
where $\textbf{r}=(x,y)$ is the position at time $t$ and $u$ is a function of time. Then
\begin{equation}
 \left|\dot{\textbf{r}}\times \ddot{\textbf{r}}\right|=x_m\cdot y_m\cdot \left|\frac{du}{dt}\right|^3
\end{equation}
where the product $x_m\cdot y_m$ is equal to the product of the square of the semi-major axis $a$, multiplied by $\sqrt{1-e^2}$ where $e$ is the eccentricity. Since the frequency locally can be approximated by $\left|\frac{du}{dt}\right|$ we get
\begin{equation}
 \omega =\left(\frac{\left|\dot{\textbf{r}}\times \ddot{\textbf{r}}\right|}{a^2\sqrt{1-e^2}}\right)^{\frac{1}{3}}
\end{equation}
Notice here that both $a$ and $e$ can be calculated by the position and velocity of the moving object (see \cite{goldstein_Book_CM_C_53}).
\section{Numerical Tests}
\subsection{The 2-body problem}
We now turn to the study of two objects interacting through a central force. The most famous example of this type, is the Kepler
problem (also called the two-body problem) that describes the motion of two bodies which attract each other. In the solar system the gravitational interaction between two bodies leads to the elliptic orbits of planets and the hyperbolic orbits of comets.

If we choose one of the bodies as the center of our coordinate system, the motion will stay in a plane. Denoting the position of
the second body by $\textbf{q}=(q_{1},q_{2})^{T}$, the Lagrangian of the system takes the form (assuming masses and gravitational constant equal to 1)
\begin{equation}
L(\textbf{q},\dot{\textbf{q}},t)=\frac{1}{2}\dot{\textbf{q}}^T \dot{\textbf{q}}+\frac{1}{|\textbf{q}|}
\end{equation}
The initial conditions are taken
\begin{equation}
 \textbf{q}=(1-\epsilon,0)^T\;,\;\dot{\textbf{q}}=\left(0,\sqrt{\frac{1+\epsilon}{1-\epsilon}}\right)^T
\end{equation}
where $\epsilon$ is the eccentricity of the orbit. In order to check the efficiency of the proposed algorithm, we shall consider only high eccentricities ($\epsilon=0.95-0.99$). In the first test, we count the number of integration steps needed for one period and for eccentricity equal to $0.95$. The results are summarized in Table \ref{table_res1}.
\begin{table}[ht]
\caption{Number of integration steps}
\centering
\begin{tabular}{ccc}
 \hline\hline \textbf{S} & \textbf{Linear} & \textbf{Phase fitted} \\
 \hline 1 & 11067  & 1789 \\
  2 & 9873   & 1124 \\
  3 & 6534   & 252  \\
  4 & 4321   & 145  \\
  5 & 3245   & 46   \\
 \hline
\end{tabular}
\label{table_res1}
\end{table}
The results have been obtained by adaptively calculating the time step, in order to keep the relative error in energy less than $10^{-6}$ (as relative error we mean the absolute error divided by the correct value).

\begin{figure}
\center
\includegraphics[scale=0.3]{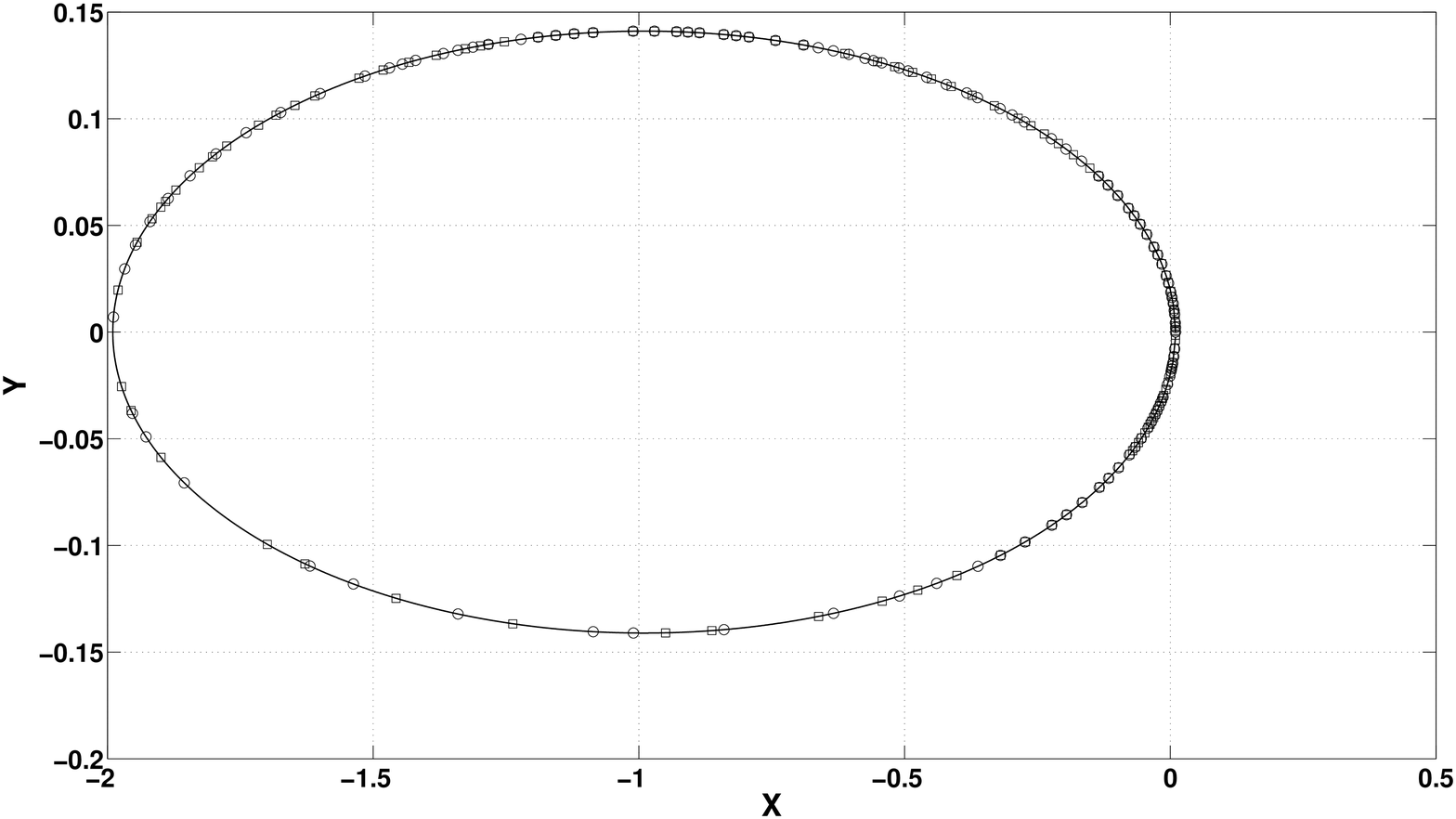}
\caption{The exact orbit for the 2-body problem for eccentricity $0.99$ for $10^6$ period (solid line), the calculated points for the first period ($\circ$) and the calculated points for the last period ($\square$).}
\label{fig_res1}
\end{figure}

\begin{figure}
\center
\includegraphics[scale=0.3]{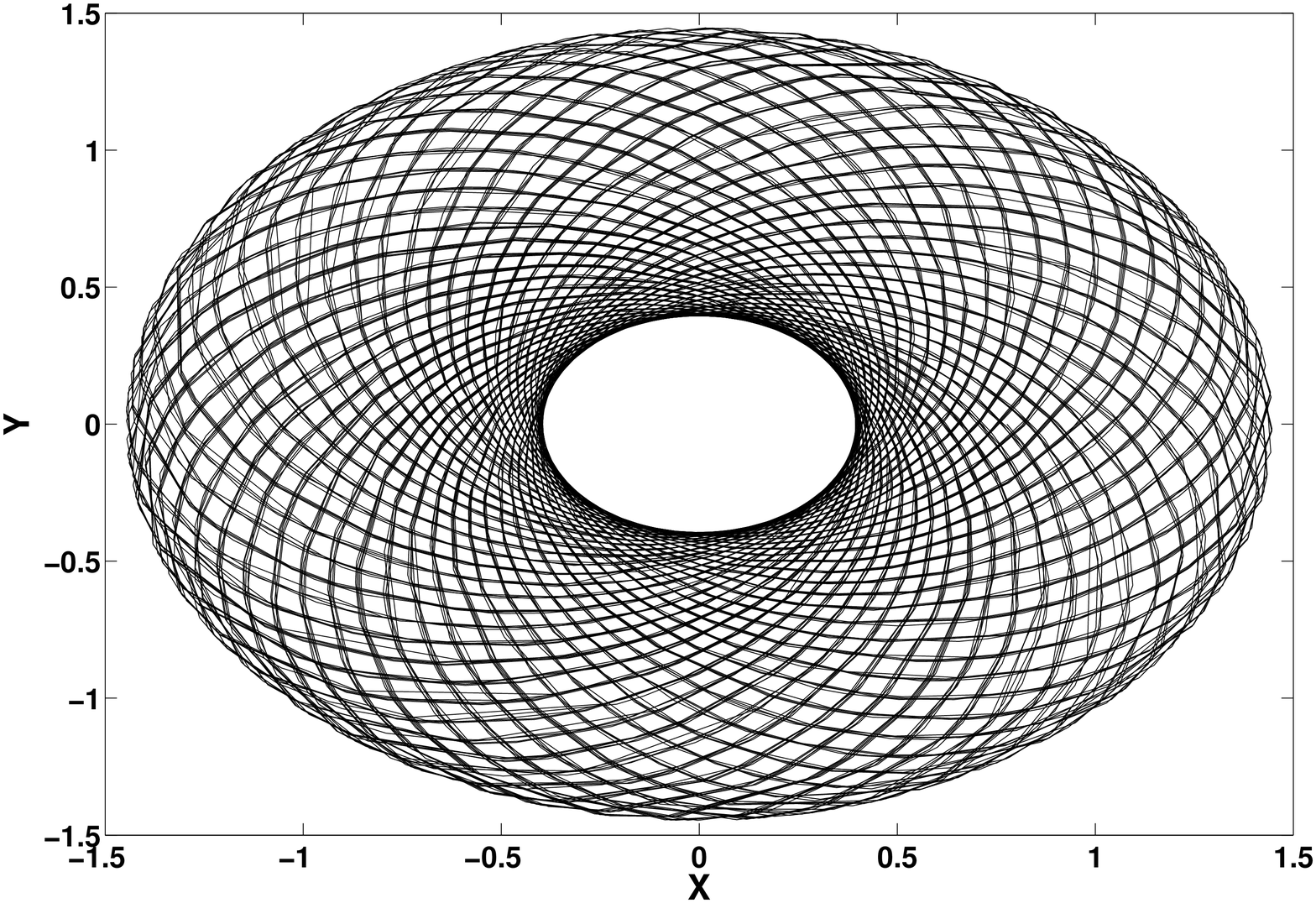}
\caption{The calculated orbit for the perturbed Kepler problem for eccentricity $0.6$ and for $10^6$ periods.}
\label{fig_res2}
\end{figure}

In the second test, we check the performance of the new method for long term integration. First, we integrate the 2-body problem for $10^6$ periods and for eccentricity $0.99$. Figure \ref{fig_res1} shows the exact orbit (solid line), the calculated points for the first period ($\circ$) and the calculated points for the last period ($\square$). Again the time step is adaptively calculated in order to keep the relative error in energy less than $10^{-6}$ (the value $10^{-6}$ of course can be changed to obtain higher or less accuracy, but in these tests has been selected because it produces $10$ full periods in less than a second in a typical personal computer). Figure \ref{fig_res2} shows the solution produced for the perturbed Kepler problem described by the Lagrangian
\begin{equation}
 L(\textbf{\textbf{q}},\dot{q})=\frac{1}{2}\dot{\textbf{q}}^T \dot{\textbf{q}}+\frac{1}{|\textbf{q}|}+\frac{0.005}{2|\textbf{q}|^3}
\end{equation}
again for $10^6$ periods and for eccentricity $0.6$ where it is clear that the solution is an ellipse that rotates slowly around one of its foci. Again the time step is adaptively control in order to keep the relative error in energy less than $10^{-6}$. All the previous test use $S=5$ as the value of intermediate points.

\subsection{Henon-Heiles Hamiltonian System}
In second test, we examine the behaviour of the new method in the Henon-Heiles Hamiltonian system. In the 1960s, a model of the motion of stars in a cylindrically symmetric, time-independent potential were investigated by astronomers (\cite{vernov_TMP_135_3_792_03}). Henon and Heiles (\cite{henon_AJ_69_73_64}) proposed the Hamiltonian
\begin{equation}
 H=\frac{1}{2}\left( \dot{x}^2 +\dot{y}^2\right) +\frac{1}{2}\left(x^2+y^2\right) +x^2y-\frac{1}{3}y^3
\end{equation}
where for small values of energy the trajectories are trivial but for higher energies, dynamic chaos emerges in the system. Setting the total energy as $E=2c^2$, Petrov produced asymptotic solutions of the form of the product of one slow and one fast oscillation with a characteristic period $T$ (see \cite{petrov_DP_52_11_635_07}). In our test, we calculate the winding number of the orbit around the origin for a half-period. The calculated values are compared with the theoretical ones produced by the asymptotic solution (\cite{petrov_DP_52_11_635_07}) and with a set of very accurate values produced in the following way: first the \textit{RKN86} $8$-stages Runge-Kutta-Nystrom pair of orders $8$ and $6$ was used (see \cite{papakostas_SIAMJSC_21_747_99}). The error tolerance of the method was set to $1E-14$ almost at the machine precision. Then, at each step, the calculated solution was projected in the manifold described by the equations $E=const$ and $J=const$ where $E,J$ the total energy and angular momentum respectively. Again the method that was applied uses $S=5$ and an adaptive time step calculation keeping the relative error in energy less than $10^{-9}$. Table \ref{table_res2} presents the results for $c=0.1$, $0.05$ and $0.0025$. Notice here that for $c=0.1$, dynamic chaos is present in the system.
\begin{table}[ht]
\caption{Winding number for half-period for the Henon-Heiles System}
\centering
\begin{tabular}{cccc}
 \hline\hline \textbf{c} & \textbf{Theoretical} & \textbf{Our Method} & \textbf{RKN86 method} \\
 \hline 0.1 & 430  & 425 & 425 \\
  0.05 & 1700  & 1697 & 1697 \\
  0.025 & 6800 & 6865 & 6865  \\
 \hline
\end{tabular}
\label{table_res2}
\end{table}

\section{Conclusions}
It has been shown in this work, that the technique of phase fitting, when it is embedded in discrete Lagrangian integrators, improves the accuracy and the efficiency of the numerical method. Following the classical application of the phase fitting technique, the discrete Lagrangian integrator is forced to solve exactly the test Lagrangian of harmonic oscillator with a given self frequency. The coefficients of the resulting integrator, depend on the frequency of the problem at each integration step. In order to improve the accuracy of the method, a set of intermediate points were added as corrections to the trigonometric path. The results show that the method can be used for long term integrations of planetary motions and this was demonstrated by applying the method to very high eccentricities ($0.99$) and for milions of periods. Moreover, a simple but quite efficient technique for frequency evaluation is proposed based on the eccentricity of the integrated orbit.

\section{Acknowledgement}
This paper is part of the 03ED51 research project, implemented within the framework of the "\emph{Reinforcement Programme of Human Research Manpower}" (\textbf{PENED}) and co-financed by National and Community Funds (25\% from the Greek Ministry of Development-General Secretariat of Research and Technology and 75\% from E.U.-European Social Fund).

% \appendix

% \section{}
% \label{}

% Bibliographic references with the natbib package:
% Parenthetical: \citep{Bai92} produces (Bailyn 1992).
% Textual: \citet{Bai95} produces Bailyn et al. (1995).
% An affix and part of a reference:
%   \citep[e.g.][Ch. 2]{Bar76}
%   produces (e.g. Barnes et al. 1976, Ch. 2).

\bibliographystyle{elsarticle-harv}
%\bibliography{../math}

\end{document}